\documentclass[twocolumn]{aastex631}
\usepackage{gensymb}

\usepackage{amsmath}
\usepackage{threeparttable}

\usepackage{graphicx,subfigure}
\usepackage{cleveref}
\crefformat{footnote}{\textsuperscript{#2#1#3}}

\begin{document}

\title{Double White Dwarf Binaries in SDSS-V DR19 : The discovery of a rare DA+DQ white dwarf binary with 31 hour orbital period}

\correspondingauthor{Gautham Adamane Pallathadka}
\email{gadaman1@jh.edu}

\author[0000-0002-5864-1332]{Gautham Adamane Pallathadka}
\affiliation{William H. Miller III Department of
Physics \& Astronomy, Johns Hopkins University, 3400 N Charles St, Baltimore, MD 21218, USA}

\author[0000-0002-0572-8012]{Vedant Chandra}
\affiliation{Center for Astrophysics $\mid$ Harvard \& Smithsonian, 60 Garden St, Cambridge, MA 02138, USA}

\author[0000-0002-2761-3005]{Boris T. G\"{a}nsicke}
\affiliation{Department of Physics, University of Warwick, Coventry CV4 7AL, UK}

\author[0000-0001-6100-6869]{Nadia L. Zakamska}
\affiliation{William H. Miller III Department of
Physics \& Astronomy, Johns Hopkins University, 3400 N Charles St, Baltimore, MD 21218, USA}

\author[0000-0002-6164-6978]{Detlev Koester}
\affiliation{Institut für Theoretische Physik und Astrophysik, University of Kiel, D-24098 Kiel, Germany}

\author[0000-0002-0632-8897]{Yossef Zenati}
\affiliation{William H. Miller III Department of
Physics \& Astronomy, Johns Hopkins University, 3400 N Charles St, Baltimore, MD 21218, USA}
\affiliation{Space Telescope Science Institute, Baltimore, MD 21218, USA}
\affiliation{Astrophysics Research Center of the Open University (ARCO), The Open University of Israel, Ra’anana 4353701, Israel}
\affiliation{Department of Natural Sciences, The Open University of Israel, Ra’anana 4353701, Israel}

\author[0000-0002-8866-4797]{Nicole R. Crumpler}
\affiliation{William H. Miller III Department of
Physics \& Astronomy, Johns Hopkins University, 3400 N Charles St, Baltimore, MD 21218, USA}

\author[0000-0002-6270-8624]{Stefan M. Arseneau}
\affiliation{Department of Astronomy \& Institute for Astrophysical Research, Boston University, 725 Commonwealth Ave., Boston, MA 02215, USA}
\affiliation{William H. Miller III Department of
Physics \& Astronomy, Johns Hopkins University, 3400 N Charles St, Baltimore, MD 21218, USA}

\author[0000-0001-5941-2286]{J. J. Hermes}
\affiliation{Department of Astronomy \& Institute for Astrophysical Research, Boston University, 725 Commonwealth Ave., Boston, MA 02215, USA}

\author[0000-0003-3903-8009]{Matthias R. Schreiber}
\affiliation{Departamento de Física, Universidad Técnica Federico Santa María, Av. España 1680, Valparaíso, Chile}

\author[0000-0002-3481-9052]{Keivan G. Stassun}
\affiliation{Department of Physics and Astronomy, Vanderbilt University, VU Station 1807, Nashville, TN 37235, USA}

\author[0000-0003-3441-9355]{Axel Schwope}
\affiliation{Leibniz-Institut fur Astrophysik Potsdam (AIP), An der Sternwarte 16, D-14482 Potsdam, Germany}

\author[0000-0002-6871-1752]{Kareem El-Badry}
\affiliation{Department of Astronomy, California Institute of Technology, 1200 East California Boulevard, Pasadena, CA 91125, USA}

\author[0000-0002-2953-7528]{Gagik Tovmassian}
\affiliation{Instituto de Astronomıa, Universidad Nacional Autónoma de México, A.P. 70-264, 04510, Mexico, D.F., México}

\author[0000-0001-7296-3533]{Tim Cunningham}
\affiliation{Center for Astrophysics | Harvard \& Smithsonian, 60 Garden St, Cambridge, MA 02138, USA}

\author[0000-0002-6770-2627]{Sean Morrison}
\affiliation{Department of Astronomy, University of Illinois at Urbana-Champaign, Urbana, IL 61801, USA}



\begin{abstract}
\noindent

Binaries of two white dwarfs (WDs) are an important class of astrophysical objects that are theorized to lead to Type Ia supernovae and are also used to gain insight into complex processes involved in stellar binary evolution. We report the discovery of SDSS~J090618.44+022311.6, a rare post-common envelope binary of a hydrogen atmospheric DA WD and a DQ WD which shows carbon absorption features, and is only the fourth such binary known. We combine the available spectroscopic, photometric, and radial velocity data to provide a self-consistent model for the binary and discuss its history as a binary DA+DQ. The system has a period of 31.17 hours with masses of 0.42 M$_{\odot}$ for DA WD and 0.49 M$_{\odot}$ for DQ WD. The corresponding cooling ages point to an Algol type of evolution with the lower mass star evolving into a DA WD first and later the massive DQ WD is formed. The system has a merger timescale of 450 Gyrs and will lead to the formation of a massive WD. With this, the number of known DA+DQ WD binaries has increased to four, and we find that their stellar properties all lie in the same range. Detailed study of more such systems is vital to understand common processes involved in the formation of this rare class of binaries and give insights towards the broader picture of WD spectral evolution. 
\end{abstract}



\section{Introduction} 
Being the remnant cores of dead stars, white dwarfs (WDs) exhibit varying core and surface properties which are determined by the progenitor star \citep{koester_atmospheric_1982,prada_moroni_very_2009,zenati_formation_2019,istrate_models_2016}. These different compositions play a crucial role during WD merger events and any ensuing supernovae \citep{pakmor_thermonuclear_2021,pakmor_fate_2022}. In most surveys, about 55-65\% of the spectroscopically identified WDs are hydrogen dominated (DA), around 25\% are featureless DC WDs, and the remaining 10\% of WDs include DB and DQ WDs that show helium absorption lines and carbon features, respectively \citep{kilic_100_2020,obrien_40_2024}. 

DQ WDs are rare: \cite{koester_carbon-rich_2019} find 572 DQ WDs among 34\,973 WDs from previous SDSS surveys, or 1.6\% of the sample. In the optical, these rare stars are characterized by the strong electronic, and ro-vibrational carbon Swan band lines due to the C$_2$ molecule, and may also have atomic C I lines or lines of ionised C II at hotter temperatures \citep{koester_carbon-rich_2019,saumon_current_2022}. The Swan band transition lines are characterized by strong band heads at 5165 \AA\ and 4737 \AA. DQ WDs can be classified into hot, warm, and cool DQ WDs based on their surface temperatures. While the formation of hot and warm DQ WDs is attributed to stellar mergers, the cool DQ WDs ($\mathrm{T_{eff}} < $10\,000 K) are formed by the dredge-up of carbon from the core through He convective zone in DB WDs \citep{koester_atmospheric_1982,camisassa_hidden_2023}. 
At temperatures lower than $\sim$12\,000 K, the helium absorption lines vanish leaving behind a DQ WD. Such DB-DQ WDs are likely to be descendents of hydrogen-defiecient post-AGB stars such as PG 1159 stars and Wolf-Rayet stars and studies of these WDs are essential to understand the physics of spectral evolution of WDs \citep{althaus_formation_2005,bedard_spectral_2020}.

Binaries of two WDs are one of the end stages of stellar binaries and allow us to gain insights into the complex processes involved in binary evolution. 
The masses of the binary stars and their orbital periods give clues to the various mass transfer processes that take place in binary evolution. These binaries enable insights into the core composition of the WDs \citep{parsons_testing_2017,zenati_formation_2019} and can help us determine the mass-radius relationship exhibited by the WDs \citep{parsons_testing_2017}. 
The number of known double WD binaries and binary candidates has ballooned due to serendipitous discoveries and dedicated surveys such as the SPY survey \citep{napiwotzki_eso_2020} and the ELM survey \citep{brown_elm_2020}. Approximately 270 WD binaries have been discovered so far \citep{munday_dbl_2024}.
The majority of these binaries are single-lined systems in which absorption lines from only one of the stars in the binary are visible, making it difficult to characterize the full system. In double-lined systems, the photospheres of both stars are visible, which makes it possible to fully characterize the binaries and infer the binary evolution pathways \citep[e.g.,][]{chandra_99-minute_2021,munday_dbl_2024}. 

The double WD binary evolution typically involves two stages of mass transfer, corresponding to each star in the binary, during which they lose the outer envelope and enter the WD stage of stellar evolution. This mass transfer can involve stable Roche lobe overflow, during which the donar star loses the envelope due to stable mass transfer, or an unstable common envelope phase during which both stars are engulfed by the envelope of the donor star \citep{woods_formation_2012,li_formation_2019}. The common envelope phase leads to orbital decay and all post common envelope binaries (PCEB) are found to be in short period orbits and unresolved \citep{li_formation_2019}.

Post-common envelope binaries of DQ WDs are rare and so far only three candidate systems have been identified. NLTT 16249 was the first such binary to be discovered, and so far the only system confirmed to be a binary through phase-resolved radial velocity (RV) follow-up \citep{vennes_117_2012,vennes_core_2012,vennes_total_2024}. NLTT 40489 (GD 184) was the second such system and was classified as a DA+DQ binary through the stellar parameter measurement \citep{kawka_spectroscopic_2006,giammichele_know_2012}. Recently, \cite{manser_desi_2024} reported the discovery of a third DA+DQ binary candidate, J0920+6926, through stellar parameter measurement. 

In this paper, we present the discovery of SDSS~J090618.44+022311.6 (hereafter SDSS~J0906+0223) the fourth DA+DQ binary candidate and only the second confirmed system with RV follow-up. We measure the stellar parameters of both DA and DQ WDs which is consistent with the dynamics of the system and known WD physics, and using the measured parameters, we deduce the formation history. The study of these rare binaries is essential to independently study the formation and physics of DQ WDs.

\section{Observations and data reduction}
\label{sec:obs}

\subsection{Spectroscopic Data}

We identified SDSS~J0906+0223 in the fifth generation of the Sloan Digital Sky Survey (SDSS-V). The target was observed as a part of the Milky Way mapper program, a multi-epoch Galactic spectroscopic program in SDSS-V (Kollmeier et al. 2025, submitted) operating at the 2.5 m telescope at the Apache Point Observatory \citep[APO,][]{gunn_25_2006} and at the Las Campanas Observatory 2.5 m telescope \citep{bowen_optical_1973} using the Baryon Oscillation Spectroscopic Survey spectrograph (BOSS; \citealt{smee_multi-object_2013}) and the Apache Point Observatory Galactic Evolution Experiment (APOGEE) spectrographs \citep{wilson_apache_2019}. The target was observed 35 times during 13.7 months between 2021-02-10 and 2022-03-25. Each spectrum covers the wavelength range from $3\,600$ \AA\ to $10\,000$ \AA\ at a resolution of R~$\sim$~1800 and has an exposure time of 900s. The data is reduced using idlspec2D version v6\_1\_12. The spectra are corrected to a heliocentric frame with absolute wavelength calibration accuracy $< 10$ $\mathrm{km~s^{-1}}$. One sub-exposure has a spurious cosmic ray line near H$\alpha$ which we mask. The system also has 4 archival SDSS observations dating back to 2001-01-15. \cite{kleinman_sdss_2013} classified the system as DA+DQ WD based on SDSS DR7 data. 

\begin{figure*}[t!]
    \centering
    \includegraphics[width=0.7\linewidth]{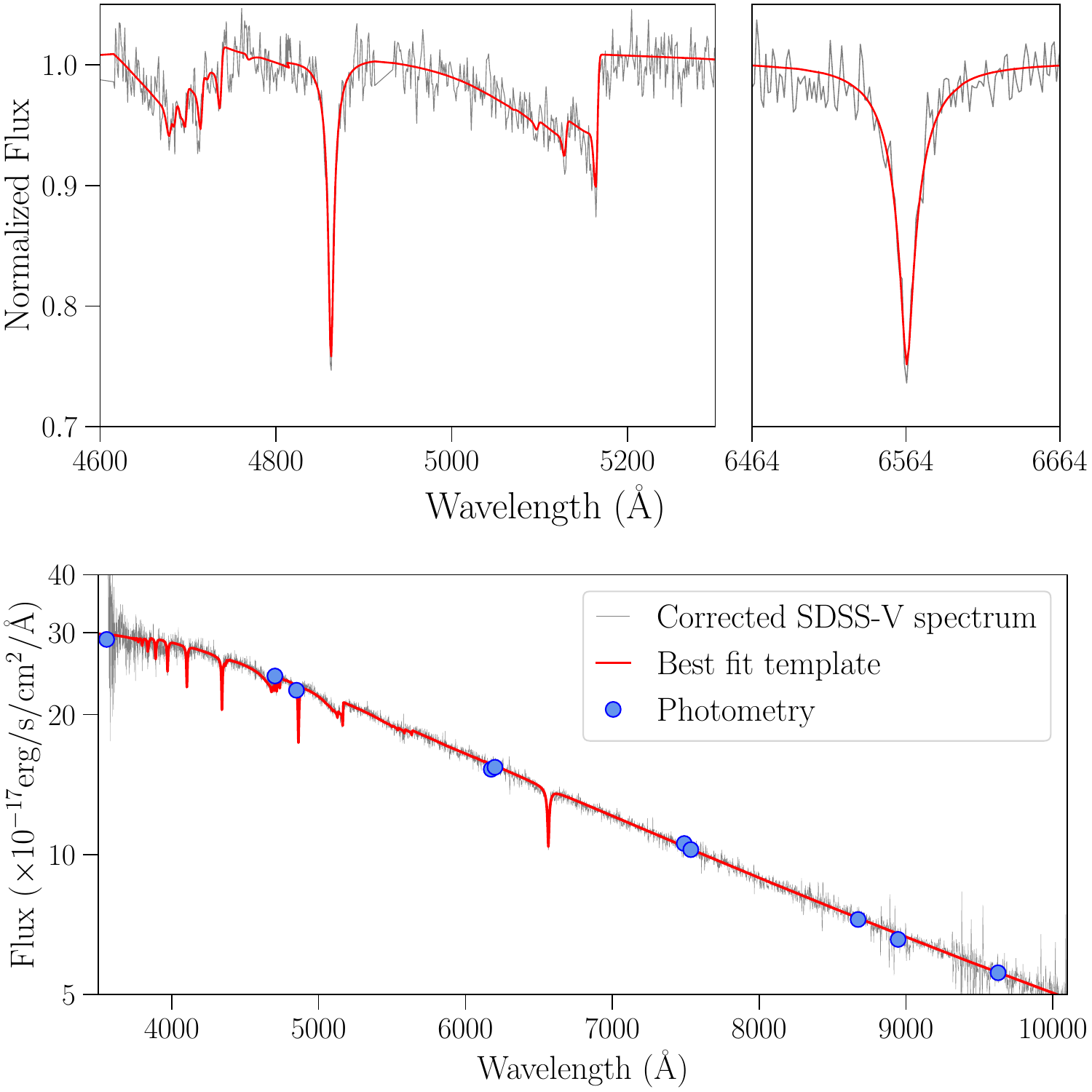}
    \caption{The best-fit spectrum in comparison with photometry and the observed SDSS spectrum. In the top plot, the comparison of model to observed spectrum is shown around H${\alpha}$ (6564 \AA), H${\beta}$ (4863 \AA), and C$_2$ Swan bands (4737 \AA\ and 5165 \AA). In the bottom plot, the best-fit spectrum is compared with photometry. In this plot, we have multiplied the observed SDSS spectrum by a constant to fix the absolute flux calibration.}
    \label{fig:spectral_fit}
\end{figure*}

The SDSS-V spectrum of the target shows the existence of the $\mathrm{C_2}$ Swan bands and the hydrogen Balmer absorption series, seen in Fig.~\ref{fig:spectral_fit}. We find that the absorption lines show radial velocity variation in anti-phase to each other, which indicated that the system is a double-lined binary of a DA WD and DQ WD.

We obtained follow-up spectra for the target using Gemini GMOS South to confirm the binary nature and obtain precise radial velocities (Program ID: GS-2022B-FT-111; PI: Gautham Adamane Pallathadka). The target was observed using B1200 grating centered at 4500 \AA\ on the night of 2023-01-19 and twice on the night of 2023-01-21. GMOS spectra covered a wavelength from 3700 \AA\ to 5300 \AA. The target was exposed for 900s per exposure for 9 total exposures with $0.75^{\prime\prime}$ slit, giving a resolution of R$\sim$2500. We reduced the data using PypeIt pipeline \citep{prochaska_pypeit_2020}, which corrected the spectra to heliocentric vacuum wavelengths. During our observation window, the amplifier \#5 of CCD2 of GMOS-S was unusable and we could not collect data for $\mathrm{C_2}$ Swan band features near 4700\AA. The spectra had flux calibration issues near the edges of the detector. We correct the red end of the Gemini spectra, beyond 4730\AA, by a second-order polynomial, which is fixed during the analysis. We also mask the spectra between 4280 \AA\ and 4320 \AA, where we find that most Gemini spectra show spectral issues not seen in SDSS-V spectra. We use the Gemini spectra only to measure the radial velocities and avoid systematic errors that may affect the stellar parameter determination due to flux calibration issues. 

\subsection{Photometric Data}

\begin{figure}
    \centering
    \includegraphics[width=0.8\columnwidth]{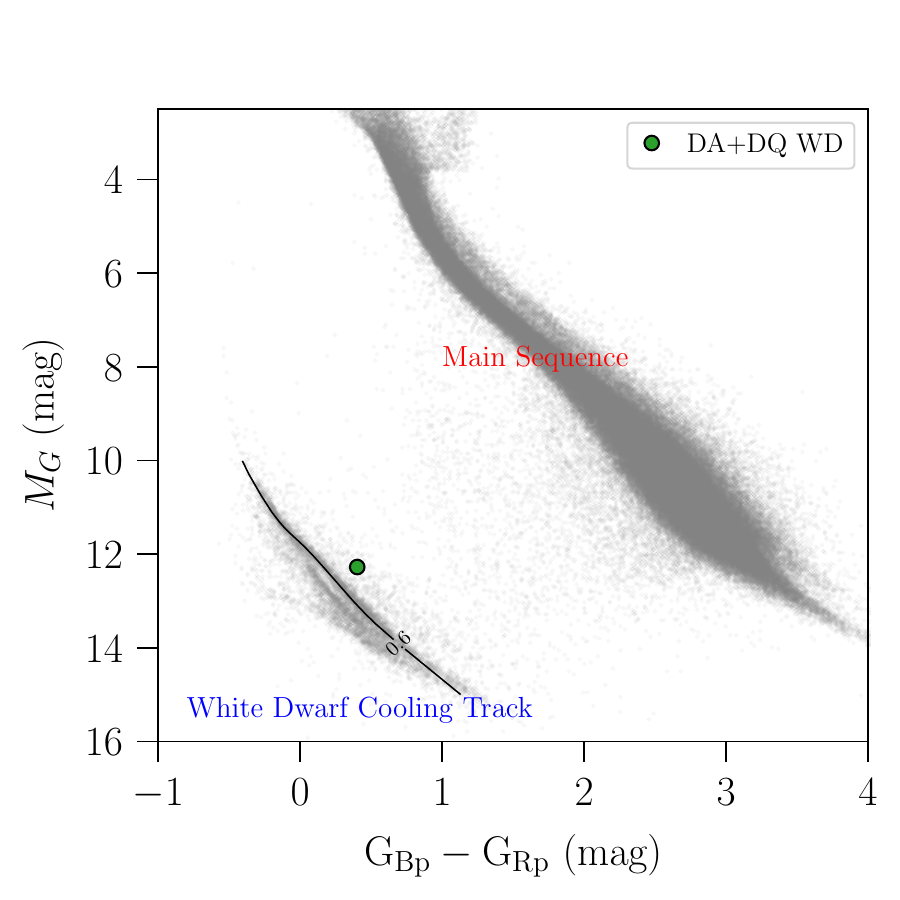}
    \caption{Color-magnitude diagram for stars within 100pc of the sun. Our target is marked in green. The sequence in the lower left is the normal WD track. We show the cooling curves for a 0.6 M$_{\odot}$ WD from \citet{bedard_spectral_2020} generated using \texttt{{WD\_models}}\textsuperscript{$\dagger$} package. Our target is overluminous compared to the cooling track of the most common 0.6 M$_{\odot}$ WD.\\ \\
    \small\textsuperscript{$\dagger$} \url{https://github.com/SihaoCheng/WD_models}}
    \label{fig:color-mag}
\end{figure}

The target SDSS~J0906+0223 has a {\it Gaia} parallax-inferred distance of $d=148 \pm 4$ pc \citep{bailer-jones_estimating_2021}. \cite{green_3d_2019} report near zero interstellar extinction along the line of sight of the target. On the {\it Gaia} color-magnitude diagram the system is over-luminous compared to the cooling track of a 0.6 M$_{\odot}$ WD, the most common WD mass (Fig.~\ref{fig:color-mag}). The target has archival photometry from GALEX NUV, SDSS {\it ugriz}, Pan-STARRS {\it grizy}, 2MASS J,H,Ks and Cat-WISE W1 and W2, covering the range from near-UV to IR. The system does not exhibit any excess in the IR. We set a floor uncertainty of 0.03 mags in all filters to take into account any systematics.

\section{Analysis}
\label{sec:analysis}

The observed spectra are modeled as a sum of net fluxes from the both stars, given by
\begin{equation}
    f_{\lambda} = \frac{\mathrm{r_{DA}}^2}{d^2} (\pi F_{\lambda,\mathrm{DA}}) + \frac{\mathrm{r_{DQ}}^2}{d^2} (\pi F_{\lambda,\mathrm{DQ}})
\end{equation}
where $\pi F_{\lambda}$ is the flux on the stellar surface and $F_{\lambda}$ is given by the model. The theoretical models assume local thermodynamic equilibrium (LTE) on the stellar atmosphere and are described in \citet{koester_accretion_2009},\citet{koester_carbon-rich_2019}, and \citet{koester_new_2020}. $\mathrm{r_{DA}}$ and $\mathrm{r_{DQ}}$ are the radii of two WDs, and $d$ is the distance to the system. A preliminary fit to the data results in DA WD mass that does not allow us to unambiguously determine the core composition, which impacts the WD mass-radius relationship. The mass of DQ WD suggested that it falls in the regime of carbon-oxygen (C/O) core WDs. Thus, we leave $\mathrm{r_{DA}}$ as a free parameter and use a C/O core with thin hydrogen layer WD cooling models by \cite{bedard_spectral_2020} for the DQ WD to obtain mass and radius from surface gravity and surface temperatures. $F_{\lambda,\mathrm{DA}}$ is determined by the surface temperature of DA WD, $\mathrm{T_{eff,DA}}$, and the surface gravity, $\log g_{\mathrm{DA}}$, and $F_{\lambda,\mathrm{DQ}}$ is determined by $\mathrm{T_{eff,DQ}}$, $\log g_{\mathrm{DQ}}$, and the carbon-to-helium abundance, $\log\mathrm{[n(C)/n(He)]}$ (hereafter, [C/He]). 
The model covers a wavelength range from 2500 \AA\ to 12\,000 \AA\ on a discrete grid of $\log g$ between 7.5 and 8.5 with spacing of 0.25, $\mathrm{T_{eff}}$ between 6000 K and 9000 K with a spacing of 250 K, and [C/He] between -5 and -8 with a spacing of 0.5. The models are interpolated using \texttt{scipy} \texttt{RegularGridInterpolator}.

DQ WDs show pressure-induced blue-shifts in their absorption lines \citep{kowalski_origin_2010,blouin_new_2019,vennes_total_2024}. To account for this shift, \cite{kowalski_origin_2010} suggested a correction to electronic transition energies of C$_2$ that is proportional to the ambient helium density with a proportionality constant $\alpha$. The DQ WD models used in this paper implement a correction similar to \cite{blouin_new_2019} with $\alpha = 0.05$, chosen such that it agrees with the sample of DQ WDs in \cite{koester_carbon-rich_2019}.

We fit the normalized fluxes around each absorption line, and the region size is chosen such that it encompasses both the absorption line and the continuum while having no overlap with the nearby absorption lines. This allows us to disentangle the out-of-phase radial velocity shift between DA and DQ. From the SDSS data we fit the Balmer lines from $\mathrm{H{\alpha}}$ through $\mathrm{H{\delta}}$ along with $\mathrm{C_{2}}$ lines at 5165 \AA\ and 4737 \AA\, while from the Gemini spectra we fit $\mathrm{H{\beta}}$, $\mathrm{H{\gamma}}$, $\mathrm{H{\delta}}$, and the $\mathrm{C_{2}}$ line at 5165 \AA.

To obtain model photometric magnitude, we convolve each model using PYPHOT\footnote{\url{https://mfouesneau.github.io/pyphot/}} in SDSS $ugriz$ and Pan-STARRS $grizy$ filters. Other photometric data lie outside the wavelength range of our model spectra and we do not use them. The model in total has 8 free parameters given by $\mathrm{T_{eff,DA}}$, $\log g_{\mathrm{DA}}$, $r_{\mathrm{DA}}$,$\mathrm{T_{eff,DQ}}$, $\log g_{\mathrm{DQ}}$, $\log\mathrm{(C/He)}$, $\mathrm{RV_{DA}}$, $\mathrm{RV_{DQ}}$, and $d$.

\subsection{Orbital parameters}
We first fit each SDSS-V exposure with the model spectra to determine tentative RVs. Using the RVs we rest-frame correct the spectra and median coadd the regions around the absorption lines. These coadded absorption lines are then simultaneously fit with photometry to determine the best-fit stellar parameters. The fit is carried out by minimizing the $\chi^2$ using lmfit \citep{newville_lmfit_2014}.
Keeping the measured parameters fixed, we remeasure the radial velocities of each SDSS-V exposure and repeat the above steps until the measured RVs between two successive runs agree within $1\sigma$, and the RV of the coadded absorption lines is within $1\sigma$ from zero.


The obtained best-fit parameters are used to measure the RVs of all the available exposures. When measuring the RVs from Gemini spectra we first fit for RVs of both stars and the three nuisance parameters associated with the second-order polynomial normalization to correct the flux calibration issue in the red end of the spectrum. We then correct the spectra by dividing out the polynomial contribution and refit them to measure the RVs. We find that the RV determination of DQ WD to be more uncertain than DA WD, along with difficulties in estimating accurate RV errors. To improve the accuracy of RVs and minimize the errors for the DQ WD, we median coadd the exposures taken within 1 hours and measure the RVs.

\begin{figure*}
\centering
\subfigure{\label{fig:a}\includegraphics[width=0.42\linewidth]{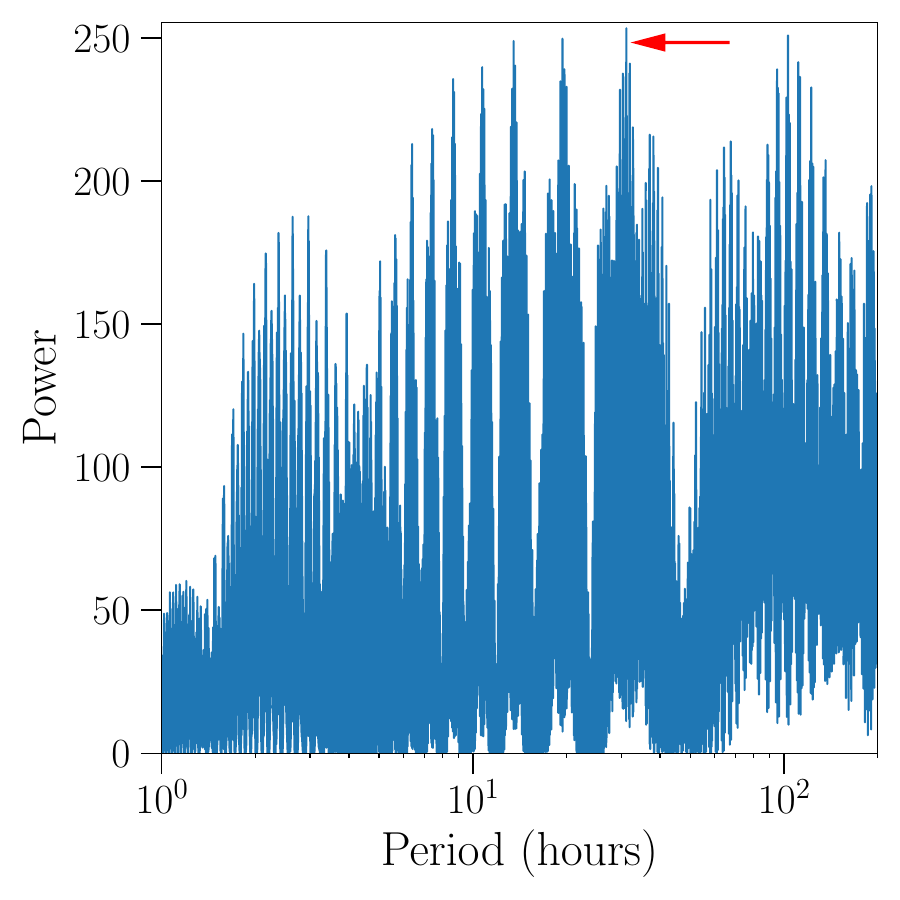}}
\subfigure{\label{fig:b}\includegraphics[width=0.42\linewidth]{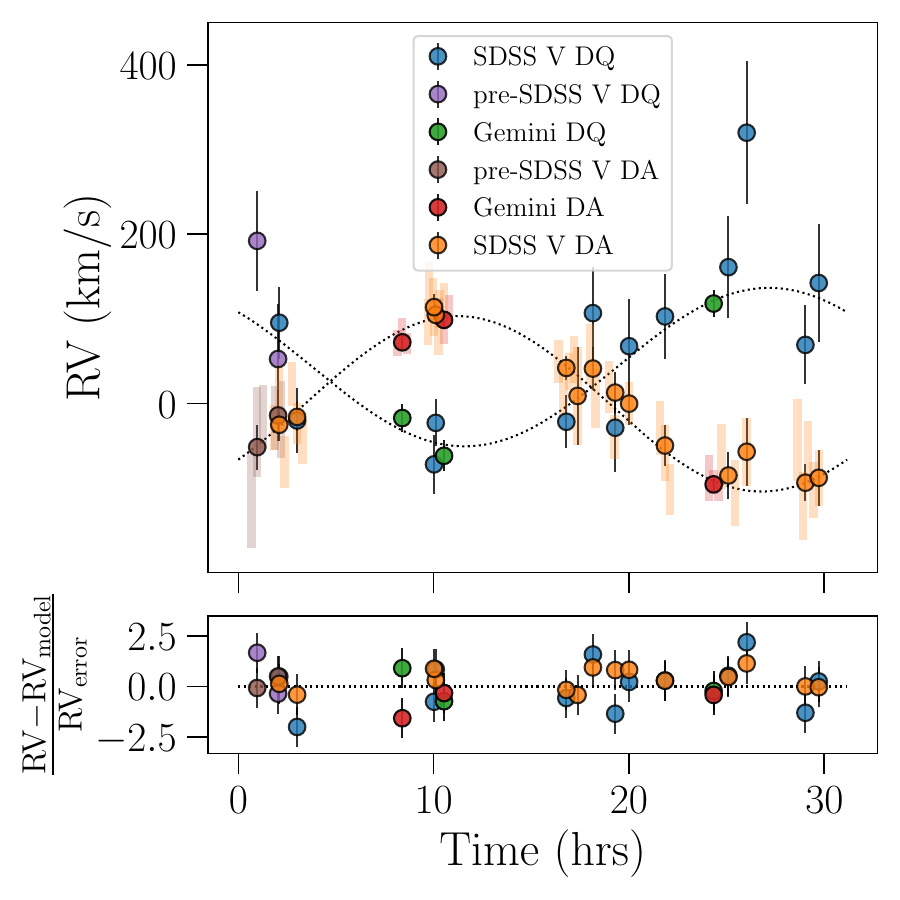}}
\caption{Left: Lomb-Scargle periodogram power as a function of orbital period is shown. The power is proportional to the $-\chi^2$ of the best-fit model at each period. The best-fit period is shown with an arrow. Right: The RV curve phase folded at the best-fit period is shown along with measured RVs from various datasets. The solid points are the RVs measured for coadded spectra, as explained in the text. For DA WD, we also show the RV of individual spectra in the background.}
\label{fig:periodogram_rv}
\end{figure*}

We discard RVs that are not well determined and have uncertainties greater than 100 $\mathrm{km~s^{-1}}$. Using the measured radial velocities for DA WD, we carry out the Lomb-Scargle periodogram in \citep{lomb_least-squares_1976,scargle_studies_1982} \texttt{astropy} (see Fig.~\ref{fig:periodogram_rv}). Using this best-fit period, we fit the RVs to a sinusoidal model and determine all the orbital parameters and their error accurately using \texttt{emcee} \citep{foreman-mackey_emcee_2013}. The sinusoidal RV model is given by 
\begin{equation}
\begin{aligned}
    \mathrm{RV_{DA}} &= \mathrm{v_{\gamma,DA} + K_{DA}\cos\left(2\pi\frac{t}{P_{orbit}} + \phi_{DA}\right)} \\
    \mathrm{RV_{DQ}} &= \mathrm{v_{\gamma,DQ} + K_{DQ}\cos\left(2\pi\frac{t}{P_{orbit}} + \phi_{DA} + \pi\right)}
\end{aligned}
\end{equation}

We find the best-fit period of 31.17 hours. The cadence of observations is insufficient to unambiguously determine the orbital period. While the periods greater than about 50 hours requires the WD masses to be much greater than what is determined using spectrophotometry, and thus ruled out, the peaks near 13 hours, 19 hours are viable solutions. The best-fit RV curve and the periodogram power spectrum is shown in Fig.~\ref{fig:periodogram_rv} and the corner plot, showing the codependence of fitted parameters, is shown in Fig.~\ref{fig:corner_all}.

To estimate the probability that we have chosen the correct period, we calculate the correctness likelihood following \cite{thorstensen_orbital_1985}. We start with the fit to observed data and calculate the contrast ratio $R_{\mathrm{obs}}$, defined as the ratio of $\chi^2$ for the best fitting period, at 31.17 hrs, and the next best fitting period, near 19 hrs. Next, we sample the RVs from the best-fit RV curve for the peak near 31.17 hrs, at the same time intervals as observed data and add normally distributed noise. For each RV, the noise is assigned from a normal distribution whose width is set by whichever is bigger between the observed RV error or the difference observed RV and the best-fit RV model. To get a conservative error estimate for the sampled RVs, we further inflate the errors by 20\%. This is repeated 1000 times and for each iteration of the sampled RVs, we carry out the Lomb-Scargle periodogram and calculate the contrast ratios $R_i$ for the best-fit period and the next best-fit period. Among these generated samples, we chose only those with contrast ratio $R_i$ within 0.1 of the observed contrast ratio, to ensure that the generated sample is similar to the observed sample. The total number of such samples is denoted by N. For this subset of the sample, we call those hits for which the best-fit period is within 0.05 hours of the observed best fit, 31.17 hrs. The ratio of hits to N gives the correctness likelihood and estimates how often the correct period is measured for a generated dataset of the same quality as the observed dataset. We find that the correctness likelihood of approximately 0.8 for the best-fit period of 31.17 hrs and conclude that there is 80\% probability that a true period is within 31.17 $\pm$ 0.05 hrs.

To ensure that the ambiguity in period determination does not affect stellar parameter measurement, we examine the fit for alternate solutions at shorter periods. The best-fit models for alternate solutions result in smaller inclinations, while leaving the RV semi-amplitude's approximately the same. We find that the stellar parameters measured in the next section largely remain unchanged for alternate period solutions and consistent with the reported uncertainties. 

The difference in masses leads to a difference of $\sim$ 6 $\mathrm{km~s^{-1}}$ in gravitational red-shifts. Taking this into account, and the uncertainties associated with determining the zero-point velocities, we find a net difference of 
(37 $\pm$ 9) $\mathrm{km~s^{-1}}$ in systemic velocities between the two stars. The pressure-shift of Swan band and the associated correction, as described in Sec.\ref{sec:analysis}, is insignificant at the temperature range we find for the DQ WD. Thus, this difference in systemic velocity is likely due to the underestimation of RV errors and inaccurate RV measurement for the DQ WD using low-SNR, low-resolution spectra.

\subsection{Stellar Parameters}
With the RV semi-amplitude of the DA white dwarf $\mathrm{K_{DA}}$ and stellar parameters of both stars, we expect the radial velocity of the DQ star to be $\mathrm{K_{DQ}} = \mathrm{K_{DA}}m_{\mathrm{DA}}/m_{\mathrm{DQ}}$, from Kepler's laws. Both $\mathrm{K_{DA}}$ and $\mathrm{K_{DQ}}$ are constrained by the orbital fit. Furthermore, $\sin i$, where $i$ is the inclination of the orbit, has to be less than 1. These place constraints on the masses of either star. We refit the coadded SDSS-V spectrum with $K_{\mathrm{DA}}$ as an additional parameter, and now include the non-uniform priors set by $K_{\mathrm{DA}}$, $K_{\mathrm{DQ}}$, and $\sin i$ as follows:

\begin{align}
\begin{split}
    \log \mathcal{L} = & -\frac{1}{2} \bigg[ \chi_{\mathrm{spec,phot}}^2 + \sum_{\mathrm{spec,phot}}\ln(2\pi\sigma^2) \bigg] \\
    & -\frac{1}{2} \sum_{\mathrm{DA,DQ}}\bigg[\left(\frac{\mathrm{K}_{i} - \mathrm{K}_{i,0}}{\sigma_{i,0}}\right)^2 + \ln(2\pi\sigma_{i,0}^2)\bigg] \\
    & -\frac{1}{2} \bigg[ \frac{{(\varpi - \varpi_{\mathrm{zp}} - 1/d_{\mathrm{pc}})^2}}{\sigma_{\varpi}^2} + \ln(2\pi\sigma_{\varpi}^2) \bigg] \\
    & - \Tilde{H}(\sin i) \\
\end{split}
\end{align}

Here, $\mathrm{K}_{i,0}$ is the best-fit parameter from the orbital fit and $\sigma_{\mathrm{i},0}$ is the associated uncertainty. The subscript `spec' corresponds to the spectroscopic fit where observed spectra are fit with model spectra and `phot' corresponds to the photometric fit where observed magnitudes are fit with the convolved model magnitudes. $\varpi$ is the \textit{Gaia} parallax in arc seconds, $\varpi_{\mathrm{zp}}$ is the parallax zero-point and equals to -0.029 mas \citep{bailer-jones_estimating_2021}, $\sigma_{\varpi}$ is the parallax error, and the $d_{\mathrm{pc}}$ is the distance to target in parsecs. $\Tilde{H}(x)$ is the modified Heaviside function that is 0 for $\sin i<1$ and $\infty$ otherwise (practically, we set it to a very large number and not $\infty$).

\begin{table}
\centering
    \caption{System parameters for SDSS~J090618.44+022311.6}
    \begin{tabular}{c c} 
         \hline
         Parameter & Value \\ \hline 
         \textit{Gaia} DR3 Source ID & 577257520277310848 \\
         R.A. (J2000) & 09:06:18.44 \\ 
         Dec. (J2000) & +02:23:11.66 \\
         \hline
         \multicolumn{2}{c}{Orbital Parameters} \\
         \hline 
         Period (hours) & 31.17 $\pm$ 0.05 \\
         $\mathrm{K_{DA}}$ ($\mathrm{km~s^{-1}}$) & 99 $\pm$ 4 \\
         $\mathrm{K_{DQ}}$ ($\mathrm{km~s^{-1}}$)$^{\dagger}$ & 85 $\pm$ 7 \\
         $\mathrm{v_{\gamma,DA}}$ ($\mathrm{km~s^{-1}}$) & 0 $\pm$ 4 \\
         $\mathrm{v_{\gamma,DQ}}$ ($\mathrm{km~s^{-1}}$) & 43 $\pm$ 8 \\
         i (deg)$^{\dagger}$ & 78 $\pm$ 6\\
         \hline 
         \multicolumn{2}{c}{Stellar Parameters} \\
         \hline 
         d (pc) & 146 $\pm$ 3 \\
         $\log g _{\mathrm{DA}}$ (cm/s$^2$) & 7.68 $\pm$ 0.06 \\
         $\log g _{\mathrm{DQ}}$ (cm/s$^2$) & 7.85 $\pm$ 0.04 \\
         $\mathrm{T_{eff,DA}}$ (K) & 6970 $\pm$ 50 \\
         $\mathrm{T_{eff,DQ}}$ (K) & 7530 $\pm$ 50 \\
         $\mathrm{[C/He]}$ & -5.56 $\pm$ 0.06 \\
         $\mathrm{r_{DA}}$ (r$_{\odot}$) & 0.0155 $\pm$ 0.0005 \\
         $\mathrm{r_{DQ}}$ (r$_{\odot}$) & 0.0138 $\pm$ 0.0003 \\
         $\mathrm{m_{DA}}$ (M$_{\odot}$) & 0.42 $\pm$ 0.05 \\
         $\mathrm{m_{DQ}}$ (M$_{\odot}$) & 0.49 $\pm$ 0.02 \\
         $\mathrm{\tau_{c,DA}}$ (Gyrs) & $\lesssim$ 2 \\
         $\mathrm{\tau_{c,DQ}}$ (Gyrs) & $\sim$ 1 \\
         \hline
    \end{tabular}
    \label{tab:res}
    \\[6pt]
    {\small $^\ast$ 1$\sigma$ errors propagated using the Monte-Carlo samples are reported.}
\end{table}

\section{Results}
\label{sec:results}
\begin{figure*}[t!]
    \centering
    \includegraphics[width=0.43\textwidth]{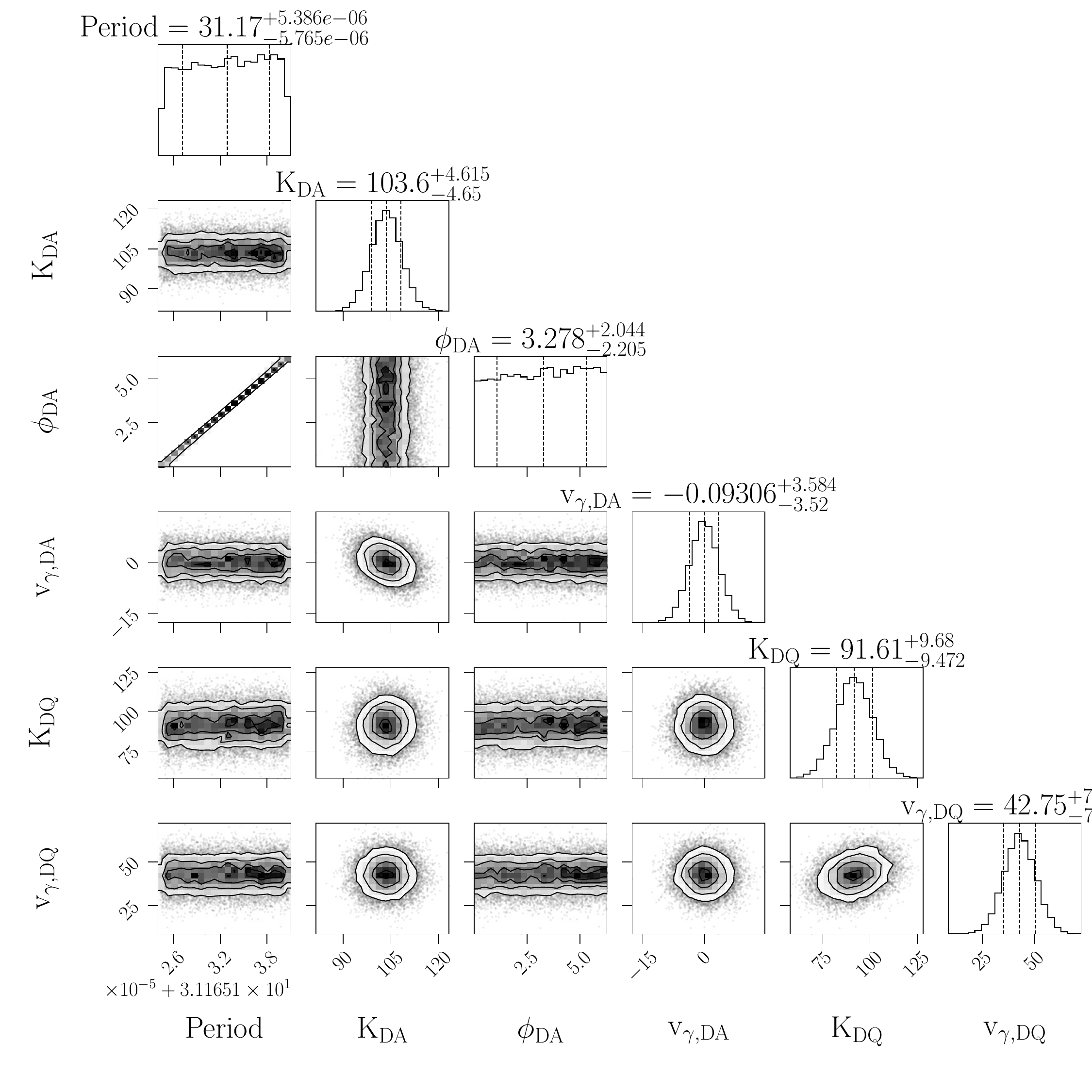}
    \includegraphics[width=0.56\textwidth]{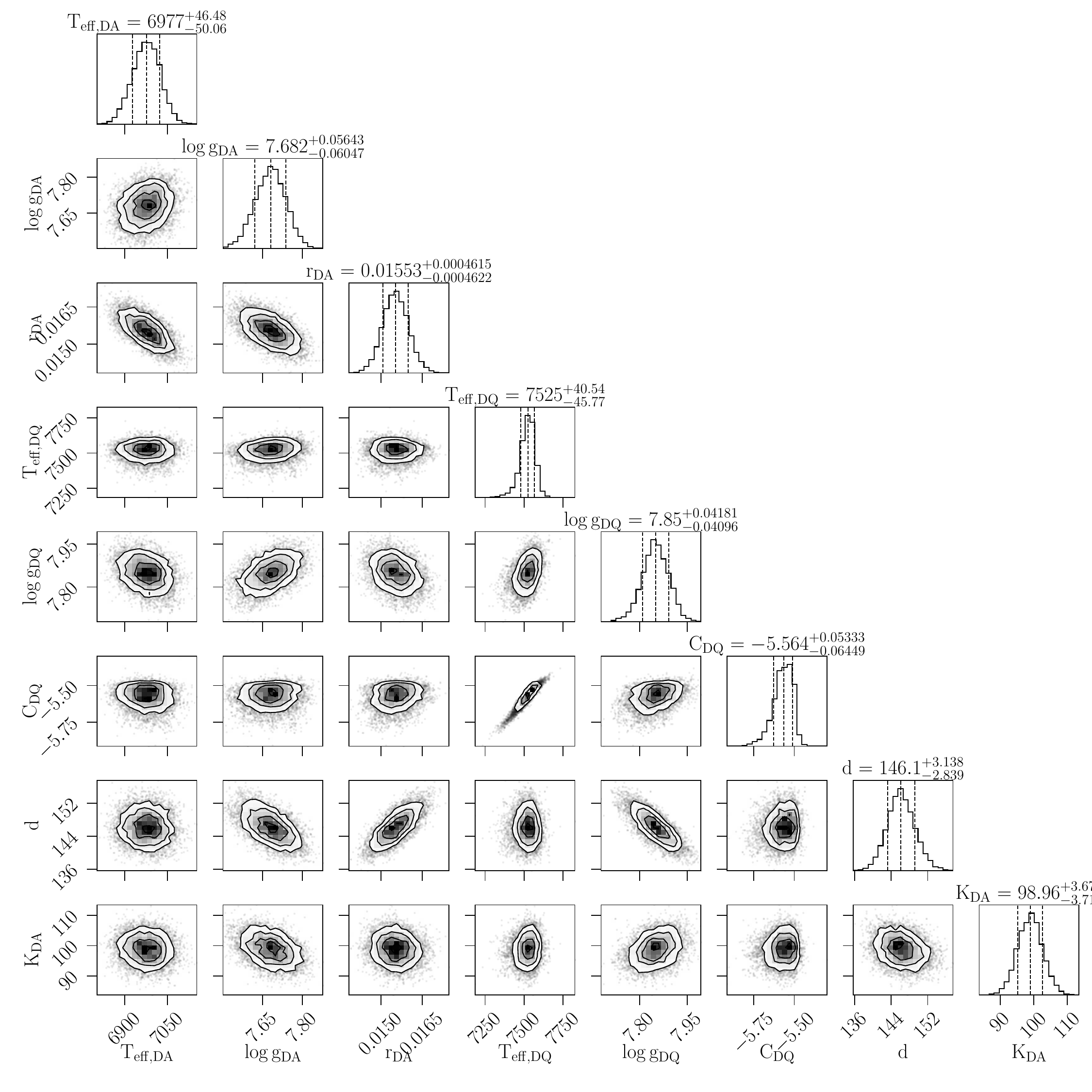}
    \caption{The results of \texttt{emcee} fit of the RVs (left) and the combined fit of spectrophotometry and the RV constraint (right).}
    \label{fig:corner_all}
\end{figure*}

The best-fit parameters are summarized in Table~\ref{tab:res} and the best-fit model is compared with observed data in Fig.~\ref{fig:spectral_fit}. The corner plot is shown in Fig.~\ref{fig:corner_all}. We find $\log g$'s of 7.68 and 7.85, and temperatures of 6977 K and 7525 K for DA and DQ WDs, respectively. Using the mass-radius relationship from \cite{bedard_spectral_2020} we find a mass of 0.49 M$_{\odot}$ for the DQ WD. For the DA WD we find a mass of 0.42 M$_{\odot}$.

Below 0.3 M$_{\odot}$ and above 0.5 M$_{\odot}$, the core of a WD can be approximated by pure-He and pure carbon-oxygen (C/O) core \citep{prada_moroni_very_2009}. In the intermediate regime, the core can have a mixed composition and be a hybrid He/C/O WD \citep{zenati_formation_2019}. The mass of the DQ WD is consistent with C/O core WD but we cannot rule out some helium in the core. The radius and mass of the DA WD are consistent with both He-core and C/O core models, given the large measurement uncertainties and is likely to have hybrid He/C/O core. 

We estimate a cooling age of $\sim$1 Gyr for the DQ WD using cooling sequences by \citet{bedard_spectral_2020}, assuming a C/O dominant core.
For the DA WD, the He-core cooling sequences by \cite{althaus_new_2013} predict a cooling age of 2.4 Gyr, and the C/O core cooling sequences by \citet{bedard_spectral_2020} predict a cooling age of 1 Gyr. The cooling age of a WD with a hybrid of He/C/O core is expected to be within this range, with the exact age determined by the fraction of He in the core. We conclude that the cooling age of DA WD is greater than the cooling age of the DQ WD and less than about 2 Gyr.

\section{Discussion}
\label{sec:discussion}

\subsection{Comparison between DA+DQ binaries}
\begin{figure}[t!]
    \centering
    \includegraphics[width=\linewidth]{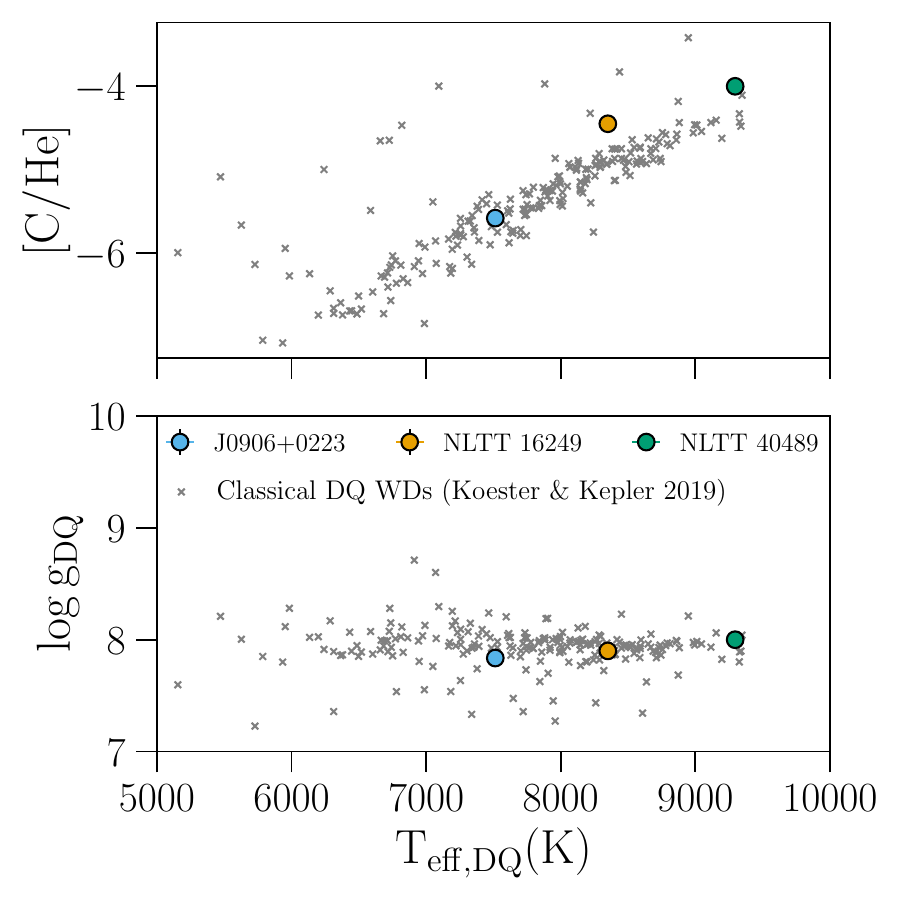}
    \caption{The stellar parameters for the three DA+DQ WD binaries are compared with the properties of single DQ WDs. All the binaries show properties consistent with single star evolution.}
    \label{fig:dadq_vs_sample}
\end{figure}

The constraints we obtain from fitting the source's spectrum and SED are consistent with the dynamical constraints of the system and therefore act as an independent test of the DQ WD physics. \cite{vennes_core_2012} find a trace amount of nitrogen on the surface of NLTT~16249, which was the first such discovery for DQ WDs. We find no traces of nitrogen in the spectra of SDSS~J0906+0223. In Fig.~\ref{fig:dadq_vs_sample} we compare the stellar properties of DQ WDs in known DA+DQ binary candidates to the sample of cool classical DQ WDs that have single star evolution. We find that the DQ WD in SDSS~J0906+0223 is consistent with single DQs and there is no evidence for the presence of any accreted material from the mass transfer phases of binary evolution.
SDSS~J0906+0223 likely originated as a DA+DB WD binary which then transitioned into DA+DBQ WD binary after 400 Myrs, and finally to pure DA+DQ after a Gyr \citep{althaus_formation_2005}. 

In Table~\ref{tab:dadq_summary} we compare the stellar parameters of known DA+DQ binaries. We find that the stellar parameters for all binaries lie in a similar range. The measured orbital periods for the two systems also lie in the same range, which could hint towards a common formation mechanism. A detailed study of the other two binary candidates and the discovery of more DA+DQ binaries is essential to characterize this class of binaries and learn about any similarities in their formation history.

\begin{table*}
\centering
\caption{The stellar parameters of known DA+DQ binaries}
\begin{tabular*}{\linewidth}{@{\extracolsep{\fill}} ccccccccc}
\hline
     DA+DQ binaries & d & Orbital Period & $\mathrm{T_{eff,DA}}$ & $\log g_{\mathrm{DA}}$ & $\mathrm{T_{eff,DQ}}$ & $\log g_{\mathrm{DQ}}$ & [C/He] & Reference \\
     & (pc) & (hrs) & (K) & & (K) & & \\
\hline
   SDSS J0906+0223 & 146 & 31.17 & 6977 & 7.68 & 7525 & 7.85 & -5.56 & This work\\
   NLTT 16249 & 57.8 & 28.17 & 8132 & 7.80 & 8347 & 7.86 & -4.45 & \cite{vennes_total_2024}\\
   NLTT 40489 & 70.0 & -- & 6800 & 8.0 & 9296 & 8.0 & -4.0 & \cite{giammichele_know_2012} \\
   J0920+6926$^{\dagger}$ & 164.5 & -- & -- & --  & 6250 & 7.33 & -6.5 & \cite{manser_desi_2024} \\
\hline
\end{tabular*}
    \\[6pt]
    {$^{\dagger}$ The stellar fits to J0920+6926 assumes that it is a single object and the measured parameters may not be accurate}
    \label{tab:dadq_summary}.
\end{table*}

\subsection{Formation history}
The formation of double WD binaries involves two phases of mass losses corresponding to the two WDs. Binary evolution calculations indicate that if both phases involve unstable mass transfer (common envelope phase) then the final orbital period can only be a few hours, whereas two stable mass transfer phases instead lead to orbital periods much larger than what we find \citep{nelemans_reconstructing_2000,li_formation_2019}. Thus, the formation of most double WD binaries with short periods involves a first phase of stable Roche lobe overflow mass transfer and the second phase of common envelope evolution \citep{nelemans_reconstructing_2000,woods_formation_2012}. The cooling ages suggest that the DA WD formed before the DQ WD, with the less massive WD being older. 

The discussion about formation history of SDSS J0906+0223 is complicated due to the uncertainty in the period. We discuss the most likely formation history for the system for the best-fit period of 31.17 hours. For orbital period of 13 hours or lower, the formation history is not as clear, and two common envelope phases cannot be immediately ruled out.

We propose an Algol-type evolution for the formation that has been discussed in detail by \citet{nelemans_reconstructing_2000} and \citet{woods_formation_2012}. The system likely started off with a binary of two main-sequence stars with masses around 1-2 M$_{\odot}$, and a mass ratio close to 1, in an orbit with a period of 10-100 days. After a few Gyrs on the main sequence, the more massive of the two ascends the asymptotic giant branch, fills the Roche lobe and, given the similar masses, begins stable mass transfer on to the companion. This leads to an increase in the mass of the companion and increased orbital separation. The stripped primary has helium in its core and hydrogen on its surface, and evolves to be a 0.42 M$_{\odot}$ DA WD with either a He core or a hybrid He/C/O core \citep{zenati_formation_2019}. 

After 1 Gyr the now more massive companion ascends the asymptotic giant branch and eventually fills its Roche lobe. The large mass ratio between the two stars leads to the common envelope phase accompanied by a decrease in the orbital separation, and the eventual common envelope ejection leads to the formation of C/O core WD with hydrogen and helium in the surface. Given that the mass of the eventually formed DQ WD, 0.49 M$_{\odot}$, falls in range of both pure C/O core WD and C/O core WD with some helium, we cannot rule out an alternative formation scenario. After the formation of DA WD, the now more massive companion instead can enter the common envelope phase at the tip of the giant branch leading to the formation of hot-subdwarf with predominantly C/O and some helium in the core. These two scenarios differ only in the core-composition of the eventual DQ WD, and we cannot rule out either scenario. This 0.49 M$_{\odot}$ WD then experiences a thermal pulse that leads to ignition of remaining hydrogen, leading to the formation of 0.49 M$_{\odot}$ helium atmospheric WD \citep{althaus_formation_2005}. In a few hundred million years, as the WD cools down, episodes of carbon dredge-up lead to the formation of DQ WD which eventually loses all helium features as the star cools below 12\,000 K.


\section*{Acknowledgments}

GAP and NLZ are thankful to JHU for support through the President’s Frontier Award to NLZ and to the Space@Hopkins Seed Grant which enabled early stages of this research program. This project has received funding from the European Research Council (ERC) under the European Union’s Horizon 2020 research and innovation programme (Grant agreement No. 101020057). N.R.C is supported by the National Science Foundation Graduate Research Fellowship Program under Grant No. DGE2139757. Any opinions, findings, and conclusions or recommendations expressed in this material are those of the author and do not necessarily reflect the views of the National Science Foundation. 

Funding for the Sloan Digital Sky Survey V has been provided by the Alfred P. Sloan Foundation, the Heising-Simons Foundation, the National Science Foundation, and the Participating Institutions. SDSS acknowledges support and resources from the Center for High-Performance Computing at the University of Utah. SDSS telescopes are located at Apache Point Observatory, funded by the Astrophysical Research Consortium and operated by New Mexico State University, and at Las Campanas Observatory, operated by the Carnegie Institution for Science. The SDSS web site is \url{www.sdss.org}.

SDSS is managed by the Astrophysical Research Consortium for the Participating Institutions of the SDSS Collaboration, including the Carnegie Institution for Science, Chilean National Time Allocation Committee (CNTAC) ratified researchers, Caltech, the Gotham Participation Group, Harvard University, Heidelberg University, The Flatiron Institute, The Johns Hopkins University, L'Ecole polytechnique f\'{e}d\'{e}rale de Lausanne (EPFL), Leibniz-Institut f\"{u}r Astrophysik Potsdam (AIP), Max-Planck-Institut f\"{u}r Astronomie (MPIA Heidelberg), Max-Planck-Institut f\"{u}r Extraterrestrische Physik (MPE), Nanjing University, National Astronomical Observatories of China (NAOC), New Mexico State University, The Ohio State University, Pennsylvania State University, Smithsonian Astrophysical Observatory, Space Telescope Science Institute (STScI), the Stellar Astrophysics Participation Group, Universidad Nacional Aut\'{o}noma de M\'{e}xico, University of Arizona, University of Colorado Boulder, University of Illinois at Urbana-Champaign, University of Toronto, University of Utah, University of Virginia, Yale University, and Yunnan University.

\software{astropy \citep{astropy_collaboration_astropy_2013,astropy_collaboration_astropy_2018,astropy_collaboration_astropy_2022}, numpy \citep{harris_array_2020}, scipy \citep{virtanen_scipy_2020}, matplotlib \citep{hunter_matplotlib_2007}}





\bibliography{references}{}

\begin{thebibliography}{}
\expandafter\ifx\csname natexlab\endcsname\relax\def\natexlab#1{#1}\fi
\providecommand{\url}[1]{\href{#1}{#1}}
\providecommand{\dodoi}[1]{doi:~\href{http://doi.org/#1}{\nolinkurl{#1}}}
\providecommand{\doeprint}[1]{\href{http://ascl.net/#1}{\nolinkurl{http://ascl.net/#1}}}
\providecommand{\doarXiv}[1]{\href{https://arxiv.org/abs/#1}{\nolinkurl{https://arxiv.org/abs/#1}}}

\bibitem[{Althaus {et~al.}(2013)Althaus, Miller~Bertolami, \&
  Córsico}]{althaus_new_2013}
Althaus, L.~G., Miller~Bertolami, M.~M., \& Córsico, A.~H. 2013, Astronomy \&
  Astrophysics, 557, A19, \dodoi{10.1051/0004-6361/201321868}

\bibitem[{Althaus {et~al.}(2005)Althaus, Serenelli, Panei, Córsico,
  García-Berro, \& Scóccola}]{althaus_formation_2005}
Althaus, L.~G., Serenelli, A.~M., Panei, J.~A., {et~al.} 2005, Astronomy \&
  Astrophysics, 435, 631, \dodoi{10.1051/0004-6361:20041965}

\bibitem[{{Astropy Collaboration} {et~al.}(2013){Astropy Collaboration},
  Robitaille, Tollerud, Greenfield, Droettboom, Bray, Aldcroft, Davis,
  Ginsburg, Price-Whelan, Kerzendorf, Conley, Crighton, Barbary, Muna,
  Ferguson, Grollier, Parikh, Nair, Unther, Deil, Woillez, Conseil, Kramer,
  Turner, Singer, Fox, Weaver, Zabalza, Edwards, Azalee~Bostroem, Burke, Casey,
  Crawford, Dencheva, Ely, Jenness, Labrie, Lim, Pierfederici, Pontzen, Ptak,
  Refsdal, Servillat, \& Streicher}]{astropy_collaboration_astropy_2013}
{Astropy Collaboration}, Robitaille, T.~P., Tollerud, E.~J., {et~al.} 2013,
  Astronomy and Astrophysics, 558, A33, \dodoi{10.1051/0004-6361/201322068}

\bibitem[{{Astropy Collaboration} {et~al.}(2018){Astropy Collaboration},
  Price-Whelan, Sipőcz, Günther, Lim, Crawford, Conseil, Shupe, Craig,
  Dencheva, Ginsburg, VanderPlas, Bradley, Pérez-Suárez, de~Val-Borro,
  Aldcroft, Cruz, Robitaille, Tollerud, Ardelean, Babej, Bach, Bachetti,
  Bakanov, Bamford, Barentsen, Barmby, Baumbach, Berry, Biscani, Boquien,
  Bostroem, Bouma, Brammer, Bray, Breytenbach, Buddelmeijer, Burke, Calderone,
  Cano~Rodríguez, Cara, Cardoso, Cheedella, Copin, Corrales, Crichton,
  D'Avella, Deil, Depagne, Dietrich, Donath, Droettboom, Earl, Erben, Fabbro,
  Ferreira, Finethy, Fox, Garrison, Gibbons, Goldstein, Gommers, Greco,
  Greenfield, Groener, Grollier, Hagen, Hirst, Homeier, Horton, Hosseinzadeh,
  Hu, Hunkeler, Ivezić, Jain, Jenness, Kanarek, Kendrew, Kern, Kerzendorf,
  Khvalko, King, Kirkby, Kulkarni, Kumar, Lee, Lenz, Littlefair, Ma, Macleod,
  Mastropietro, McCully, Montagnac, Morris, Mueller, Mumford, Muna, Murphy,
  Nelson, Nguyen, Ninan, Nöthe, Ogaz, Oh, Parejko, Parley, Pascual, Patil,
  Patil, Plunkett, Prochaska, Rastogi, Reddy~Janga, Sabater, Sakurikar,
  Seifert, Sherbert, Sherwood-Taylor, Shih, Sick, Silbiger, Singanamalla,
  Singer, Sladen, Sooley, Sornarajah, Streicher, Teuben, Thomas, Tremblay,
  Turner, Terrón, van Kerkwijk, de~la Vega, Watkins, Weaver, Whitmore,
  Woillez, Zabalza, \& {Astropy
  Contributors}}]{astropy_collaboration_astropy_2018}
{Astropy Collaboration}, Price-Whelan, A.~M., Sipőcz, B.~M., {et~al.} 2018,
  The Astronomical Journal, 156, 123, \dodoi{10.3847/1538-3881/aabc4f}

\bibitem[{{Astropy Collaboration} {et~al.}(2022){Astropy Collaboration},
  Price-Whelan, Lim, Earl, Starkman, Bradley, Shupe, Patil, Corrales, Brasseur,
  Nöthe, Donath, Tollerud, Morris, Ginsburg, Vaher, Weaver, Tocknell,
  Jamieson, van Kerkwijk, Robitaille, Merry, Bachetti, Günther, Aldcroft,
  Alvarado-Montes, Archibald, Bódi, Bapat, Barentsen, Bazán, Biswas, Boquien,
  Burke, Cara, Cara, Conroy, Conseil, Craig, Cross, Cruz, D'Eugenio, Dencheva,
  Devillepoix, Dietrich, Eigenbrot, Erben, Ferreira, Foreman-Mackey, Fox,
  Freij, Garg, Geda, Glattly, Gondhalekar, Gordon, Grant, Greenfield, Groener,
  Guest, Gurovich, Handberg, Hart, Hatfield-Dodds, Homeier, Hosseinzadeh,
  Jenness, Jones, Joseph, Kalmbach, Karamehmetoglu, Kałuszyński, Kelley,
  Kern, Kerzendorf, Koch, Kulumani, Lee, Ly, Ma, MacBride, Maljaars, Muna,
  Murphy, Norman, O'Steen, Oman, Pacifici, Pascual, Pascual-Granado, Patil,
  Perren, Pickering, Rastogi, Roulston, Ryan, Rykoff, Sabater, Sakurikar,
  Salgado, Sanghi, Saunders, Savchenko, Schwardt, Seifert-Eckert, Shih, Jain,
  Shukla, Sick, Simpson, Singanamalla, Singer, Singhal, Sinha, Sipőcz,
  Spitler, Stansby, Streicher, Šumak, Swinbank, Taranu, Tewary, Tremblay,
  de~Val-Borro, Van~Kooten, Vasović, Verma, de~Miranda~Cardoso, Williams,
  Wilson, Winkel, Wood-Vasey, Xue, Yoachim, Zhang, Zonca, \& {Astropy Project
  Contributors}}]{astropy_collaboration_astropy_2022}
{Astropy Collaboration}, Price-Whelan, A.~M., Lim, P.~L., {et~al.} 2022, The
  Astrophysical Journal, 935, 167, \dodoi{10.3847/1538-4357/ac7c74}

\bibitem[{Bailer-Jones {et~al.}(2021)Bailer-Jones, Rybizki, Fouesneau,
  Demleitner, \& Andrae}]{bailer-jones_estimating_2021}
Bailer-Jones, C. A.~L., Rybizki, J., Fouesneau, M., Demleitner, M., \& Andrae,
  R. 2021, The Astronomical Journal, 161, 147, \dodoi{10.3847/1538-3881/abd806}

\bibitem[{Blouin {et~al.}(2019)Blouin, Dufour, Thibeault, \&
  Allard}]{blouin_new_2019}
Blouin, S., Dufour, P., Thibeault, C., \& Allard, N.~F. 2019, The Astrophysical
  Journal, 878, 63, \dodoi{10.3847/1538-4357/ab1f82}

\bibitem[{Bowen \& Vaughan(1973)}]{bowen_optical_1973}
Bowen, I.~S., \& Vaughan, A.~H. 1973, Applied Optics, 12, 1430,
  \dodoi{10.1364/AO.12.001430}

\bibitem[{Brown {et~al.}(2020)Brown, Kilic, Kosakowski, Andrews, Heinke,
  Agüeros, Camilo, Gianninas, Hermes, \& Kenyon}]{brown_elm_2020}
Brown, W.~R., Kilic, M., Kosakowski, A., {et~al.} 2020, The Astrophysical
  Journal, 889, 49, \dodoi{10.3847/1538-4357/ab63cd}

\bibitem[{Bédard {et~al.}(2020)Bédard, Bergeron, Brassard, \&
  Fontaine}]{bedard_spectral_2020}
Bédard, A., Bergeron, P., Brassard, P., \& Fontaine, G. 2020, The
  Astrophysical Journal, 901, 93, \dodoi{10.3847/1538-4357/abafbe}

\bibitem[{Camisassa {et~al.}(2023)Camisassa, Torres, Hollands, Koester, Raddi,
  Althaus, \& Rebassa-Mansergas}]{camisassa_hidden_2023}
Camisassa, M., Torres, S., Hollands, M., {et~al.} 2023, Astronomy \&
  Astrophysics, 674, A213, \dodoi{10.1051/0004-6361/202346628}

\bibitem[{Chandra {et~al.}(2021)Chandra, Hwang, Zakamska, Gaensicke, Hermes,
  Schwope, Badenes, Tovmassian, Bauer, Maoz, Schreiber, Toloza, Inight, Rix, \&
  Brown}]{chandra_99-minute_2021}
Chandra, V., Hwang, H.-C., Zakamska, N.~L., {et~al.} 2021, The Astrophysical
  Journal, 921, 160, \dodoi{10.3847/1538-4357/ac2145}

\bibitem[{Foreman-Mackey {et~al.}(2013)Foreman-Mackey, Hogg, Lang, \&
  Goodman}]{foreman-mackey_emcee_2013}
Foreman-Mackey, D., Hogg, D.~W., Lang, D., \& Goodman, J. 2013, Publications of
  the Astronomical Society of the Pacific, 125, 306, \dodoi{10.1086/670067}

\bibitem[{Giammichele {et~al.}(2012)Giammichele, Bergeron, \&
  Dufour}]{giammichele_know_2012}
Giammichele, N., Bergeron, P., \& Dufour, P. 2012, The Astrophysical Journal
  Supplement Series, 199, 29, \dodoi{10.1088/0067-0049/199/2/29}

\bibitem[{Green {et~al.}(2019)Green, Schlafly, Zucker, Speagle, \&
  Finkbeiner}]{green_3d_2019}
Green, G.~M., Schlafly, E., Zucker, C., Speagle, J.~S., \& Finkbeiner, D. 2019,
  The Astrophysical Journal, 887, 93, \dodoi{10.3847/1538-4357/ab5362}

\bibitem[{Gunn {et~al.}(2006)Gunn, Siegmund, Mannery, Owen, Hull, Leger, Carey,
  Knapp, York, Boroski, Kent, Lupton, Rockosi, Evans, Waddell, Anderson, Annis,
  Barentine, Bartoszek, Bastian, Bracker, Brewington, Briegel, Brinkmann,
  Brown, Carr, Czarapata, Drennan, Dombeck, Federwitz, Gillespie, Gonzales,
  Hansen, Harvanek, Hayes, Jordan, Kinney, Klaene, Kleinman, Kron, Kresinski,
  Lee, Limmongkol, Lindenmeyer, Long, Loomis, McGehee, Mantsch, Neilsen,
  Neswold, Newman, Nitta, Peoples, Pier, Prieto, Prosapio, Rivetta, Schneider,
  Snedden, \& Wang}]{gunn_25_2006}
Gunn, J.~E., Siegmund, W.~A., Mannery, E.~J., {et~al.} 2006, The Astronomical
  Journal, 131, 2332, \dodoi{10.1086/500975}

\bibitem[{Harris {et~al.}(2020)Harris, Millman, Van Der~Walt, Gommers,
  Virtanen, Cournapeau, Wieser, Taylor, Berg, Smith, Kern, Picus, Hoyer,
  Van~Kerkwijk, Brett, Haldane, Del~Río, Wiebe, Peterson, Gérard-Marchant,
  Sheppard, Reddy, Weckesser, Abbasi, Gohlke, \& Oliphant}]{harris_array_2020}
Harris, C.~R., Millman, K.~J., Van Der~Walt, S.~J., {et~al.} 2020, Nature, 585,
  357, \dodoi{10.1038/s41586-020-2649-2}

\bibitem[{Hunter(2007)}]{hunter_matplotlib_2007}
Hunter, J.~D. 2007, Computing in Science \& Engineering, 9, 90,
  \dodoi{10.1109/MCSE.2007.55}

\bibitem[{Istrate {et~al.}(2016)Istrate, Marchant, Tauris, Langer, Stancliffe,
  \& Grassitelli}]{istrate_models_2016}
Istrate, A.~G., Marchant, P., Tauris, T.~M., {et~al.} 2016, Astronomy \&
  Astrophysics, 595, A35, \dodoi{10.1051/0004-6361/201628874}

\bibitem[{Kawka \& Vennes(2006)}]{kawka_spectroscopic_2006}
Kawka, A., \& Vennes, S. 2006, The Astrophysical Journal, 643, 402,
  \dodoi{10.1086/501451}

\bibitem[{Kilic {et~al.}(2020)Kilic, Bergeron, Kosakowski, Brown, Agüeros, \&
  Blouin}]{kilic_100_2020}
Kilic, M., Bergeron, P., Kosakowski, A., {et~al.} 2020, The Astrophysical
  Journal, 898, 84, \dodoi{10.3847/1538-4357/ab9b8d}

\bibitem[{Kleinman {et~al.}(2013)Kleinman, Kepler, Koester, Pelisoli, Peçanha,
  Nitta, Costa, Krzesinski, Dufour, Lachapelle, Bergeron, Yip, Harris,
  Eisenstein, Althaus, \& Córsico}]{kleinman_sdss_2013}
Kleinman, S.~J., Kepler, S.~O., Koester, D., {et~al.} 2013, The Astrophysical
  Journal Supplement Series, 204, 5, \dodoi{10.1088/0067-0049/204/1/5}

\bibitem[{Koester(2009)}]{koester_accretion_2009}
Koester, D. 2009, Astronomy \& Astrophysics, 498, 517,
  \dodoi{10.1051/0004-6361/200811468}

\bibitem[{Koester \& Kepler(2019)}]{koester_carbon-rich_2019}
Koester, D., \& Kepler, S.~O. 2019, Astronomy \& Astrophysics, 628, A102,
  \dodoi{10.1051/0004-6361/201935946}

\bibitem[{Koester {et~al.}(2020)Koester, Kepler, \& Irwin}]{koester_new_2020}
Koester, D., Kepler, S.~O., \& Irwin, A.~W. 2020, Astronomy \& Astrophysics,
  635, A103, \dodoi{10.1051/0004-6361/202037530}

\bibitem[{Koester {et~al.}(1982)Koester, Weidemann, \&
  Zeidler}]{koester_atmospheric_1982}
Koester, D., Weidemann, V., \& Zeidler, E.~M. 1982, Astronomy and Astrophysics,
  116, 147.
\newblock \url{https://ui.adsabs.harvard.edu/abs/1982A&A...116..147K}

\bibitem[{Kowalski(2010)}]{kowalski_origin_2010}
Kowalski, P.~M. 2010, Astronomy and Astrophysics, 519, L8,
  \dodoi{10.1051/0004-6361/201015238}

\bibitem[{Li {et~al.}(2019)Li, Chen, Chen, \& Han}]{li_formation_2019}
Li, Z., Chen, X., Chen, H.-L., \& Han, Z. 2019, The Astrophysical Journal, 871,
  148, \dodoi{10.3847/1538-4357/aaf9a1}

\bibitem[{Lomb(1976)}]{lomb_least-squares_1976}
Lomb, N.~R. 1976, Astrophysics and Space Science, 39, 447,
  \dodoi{10.1007/BF00648343}

\bibitem[{Manser {et~al.}(2024)Manser, Izquierdo, Gänsicke, Swan, Koester,
  Robert, Xu, Inight, Amroota, Fusillo, Koposov, Kim, Dey, Prieto, Aguilar,
  Ahlen, Blum, Brooks, Claybaugh, Cooper, Dawson, de la Macorra, Doel,
  Forero-Romero, Gaztañaga, Gontcho, Honscheid, Kisner, Kremin, Lambert,
  Landriau, Le Guillou, Levi, Li, Meisner, Miquel, Moustakas, Nie,
  Palanque-Delabrouille, Percival, Poppett, Prada, Rezaie, Rossi, Sanchez,
  Schlafly, Schlegel, Schubnell, Seo, Silber, Tarlé, Weaver, Zhou, \&
  Zou}]{manser_desi_2024}
Manser, C.~J., Izquierdo, P., Gänsicke, B.~T., {et~al.} 2024, Monthly Notices
  of the Royal Astronomical Society, 535, 254, \dodoi{10.1093/mnras/stae2205}

\bibitem[{Munday {et~al.}(2024)Munday, Pelisoli, Tremblay, Marsh, Nelemans,
  Bédard, Toonen, Breedt, Cunningham, O’Brien, \& Dawson}]{munday_dbl_2024}
Munday, J., Pelisoli, I., Tremblay, P.~E., {et~al.} 2024, Monthly Notices of
  the Royal Astronomical Society, 532, 2534, \dodoi{10.1093/mnras/stae1645}

\bibitem[{Napiwotzki {et~al.}(2020)Napiwotzki, Karl, Lisker, Catalán,
  Drechsel, Heber, Homeier, Koester, Leibundgut, Marsh, Moehler, Nelemans,
  Reimers, Renzini, Ströer, \& Yungelson}]{napiwotzki_eso_2020}
Napiwotzki, R., Karl, C.~A., Lisker, T., {et~al.} 2020, Astronomy \&
  Astrophysics, 638, A131, \dodoi{10.1051/0004-6361/201629648}

\bibitem[{Nelemans {et~al.}(2000)Nelemans, Verbunt, Yungelson, \&
  Portegies~Zwart}]{nelemans_reconstructing_2000}
Nelemans, G., Verbunt, F., Yungelson, L.~R., \& Portegies~Zwart, S.~F. 2000,
  Astronomy and Astrophysics, 360, 1011,
  \dodoi{10.48550/arXiv.astro-ph/0006216}

\bibitem[{Newville {et~al.}(2014)Newville, Stensitzki, Allen, \&
  Ingargiola}]{newville_lmfit_2014}
Newville, M., Stensitzki, T., Allen, D.~B., \& Ingargiola, A. 2014, {LMFIT}:
  {Non}-{Linear} {Least}-{Square} {Minimization} and {Curve}-{Fitting} for
  {Python},  Zenodo, \dodoi{10.5281/ZENODO.11813}

\bibitem[{O'Brien {et~al.}(2024)O'Brien, Tremblay, Klein, Koester, Melis,
  Bédard, Cukanovaite, Cunningham, Doyle, Gänsicke, Gentile~Fusillo,
  Hollands, McCleery, Pelisoli, Toonen, Weinberger, \&
  Zuckerman}]{obrien_40_2024}
O'Brien, M.~W., Tremblay, P.~E., Klein, B.~L., {et~al.} 2024, Monthly Notices
  of the Royal Astronomical Society, 527, 8687, \dodoi{10.1093/mnras/stad3773}

\bibitem[{Pakmor {et~al.}(2021)Pakmor, Zenati, Perets, \&
  Toonen}]{pakmor_thermonuclear_2021}
Pakmor, R., Zenati, Y., Perets, H.~B., \& Toonen, S. 2021, Monthly Notices of
  the Royal Astronomical Society, 503, 4734, \dodoi{10.1093/mnras/stab686}

\bibitem[{Pakmor {et~al.}(2022)Pakmor, Callan, Collins, de Mink, Holas,
  Kerzendorf, Kromer, Neunteufel, O’Brien, Röpke, Ruiter, Seitenzahl,
  Shingles, Sim, \& Taubenberger}]{pakmor_fate_2022}
Pakmor, R., Callan, F.~P., Collins, C.~E., {et~al.} 2022, Monthly Notices of
  the Royal Astronomical Society, 517, 5260, \dodoi{10.1093/mnras/stac3107}

\bibitem[{Parsons {et~al.}(2017)Parsons, Gänsicke, Marsh, Ashley, Bours,
  Breedt, Burleigh, Copperwheat, Dhillon, Green, Hardy, Hermes, Irawati, Kerry,
  Littlefair, McAllister, Rattanasoon, Rebassa-Mansergas, Sahman, \&
  Schreiber}]{parsons_testing_2017}
Parsons, S.~G., Gänsicke, B.~T., Marsh, T.~R., {et~al.} 2017, Monthly Notices
  of the Royal Astronomical Society, 470, 4473, \dodoi{10.1093/mnras/stx1522}

\bibitem[{Prada~Moroni \& Straniero(2009)}]{prada_moroni_very_2009}
Prada~Moroni, P.~G., \& Straniero, O. 2009, Astronomy \& Astrophysics, 507,
  1575, \dodoi{10.1051/0004-6361/200912847}

\bibitem[{Prochaska {et~al.}(2020)Prochaska, Hennawi, Westfall, Cooke, Wang,
  Hsyu, Davies, Farina, \& Pelliccia}]{prochaska_pypeit_2020}
Prochaska, J., Hennawi, J., Westfall, K., {et~al.} 2020, Journal of Open Source
  Software, 5, 2308, \dodoi{10.21105/joss.02308}

\bibitem[{Saumon {et~al.}(2022)Saumon, Blouin, \&
  Tremblay}]{saumon_current_2022}
Saumon, D., Blouin, S., \& Tremblay, P.-E. 2022, Physics Reports, 988, 1,
  \dodoi{10.1016/j.physrep.2022.09.001}

\bibitem[{Scargle(1982)}]{scargle_studies_1982}
Scargle, J.~D. 1982, The Astrophysical Journal, 263, 835,
  \dodoi{10.1086/160554}

\bibitem[{Smee {et~al.}(2013)Smee, Gunn, Uomoto, Roe, Schlegel, Rockosi, Carr,
  Leger, Dawson, Olmstead, Brinkmann, Owen, Barkhouser, Honscheid, Harding,
  Long, Lupton, Loomis, Anderson, Annis, Bernardi, Bhardwaj, Bizyaev, Bolton,
  Brewington, Briggs, Burles, Burns, Castander, Connolly, Davenport, Ebelke,
  Epps, Feldman, Friedman, Frieman, Heckman, Hull, Knapp, Lawrence, Loveday,
  Mannery, Malanushenko, Malanushenko, Merrelli, Muna, Newman, Nichol, Oravetz,
  Pan, Pope, Ricketts, Shelden, Sandford, Siegmund, Simmons, Smith, Snedden,
  Schneider, SubbaRao, Tremonti, Waddell, \& York}]{smee_multi-object_2013}
Smee, S.~A., Gunn, J.~E., Uomoto, A., {et~al.} 2013, The Astronomical Journal,
  146, 32, \dodoi{10.1088/0004-6256/146/2/32}

\bibitem[{Thorstensen \& Freed(1985)}]{thorstensen_orbital_1985}
Thorstensen, J.~R., \& Freed, I.~W. 1985, The Astronomical Journal, 90, 2082,
  \dodoi{10.1086/113916}

\bibitem[{Vennes \& Kawka(2012)}]{vennes_core_2012}
Vennes, S., \& Kawka, A. 2012, The Astrophysical Journal, 745, L12,
  \dodoi{10.1088/2041-8205/745/1/L12}

\bibitem[{Vennes \& Kawka(2024)}]{vennes_total_2024}
---. 2024, The total mass of the close, double degenerate ({DA}+{DQ}) system
  {NLTT}{\textasciitilde}16249,  arXiv, \dodoi{10.48550/arXiv.2412.03144}

\bibitem[{Vennes {et~al.}(2012)Vennes, Kawka, O’Toole, \&
  Thorstensen}]{vennes_117_2012}
Vennes, S., Kawka, A., O’Toole, S.~J., \& Thorstensen, J.~R. 2012, The
  Astrophysical Journal, 756, L5, \dodoi{10.1088/2041-8205/756/1/L5}

\bibitem[{Virtanen {et~al.}(2020)Virtanen, Gommers, Oliphant, Haberland, Reddy,
  Cournapeau, Burovski, Peterson, Weckesser, Bright, Van Der~Walt, Brett,
  Wilson, Millman, Mayorov, Nelson, Jones, Kern, Larson, Carey, Polat, Feng,
  Moore, VanderPlas, Laxalde, Perktold, Cimrman, Henriksen, Quintero, Harris,
  Archibald, Ribeiro, Pedregosa, Van~Mulbregt, {SciPy 1.0 Contributors},
  Vijaykumar, Bardelli, Rothberg, Hilboll, Kloeckner, Scopatz, Lee, Rokem,
  Woods, Fulton, Masson, Häggström, Fitzgerald, Nicholson, Hagen, Pasechnik,
  Olivetti, Martin, Wieser, Silva, Lenders, Wilhelm, Young, Price, Ingold,
  Allen, Lee, Audren, Probst, Dietrich, Silterra, Webber, Slavič, Nothman,
  Buchner, Kulick, Schönberger, De~Miranda~Cardoso, Reimer, Harrington,
  Rodríguez, Nunez-Iglesias, Kuczynski, Tritz, Thoma, Newville, Kümmerer,
  Bolingbroke, Tartre, Pak, Smith, Nowaczyk, Shebanov, Pavlyk, Brodtkorb, Lee,
  McGibbon, Feldbauer, Lewis, Tygier, Sievert, Vigna, Peterson, More, Pudlik,
  Oshima, Pingel, Robitaille, Spura, Jones, Cera, Leslie, Zito, Krauss,
  Upadhyay, Halchenko, \& Vázquez-Baeza}]{virtanen_scipy_2020}
Virtanen, P., Gommers, R., Oliphant, T.~E., {et~al.} 2020, Nature Methods, 17,
  261, \dodoi{10.1038/s41592-019-0686-2}

\bibitem[{Wilson {et~al.}(2019)Wilson, Hearty, Skrutskie, Majewski, Holtzman,
  Eisenstein, Gunn, Blank, Henderson, Smee, Nelson, Nidever, Arns, Barkhouser,
  Barr, Beland, Bershady, Blanton, Brunner, Burton, Carey, Carr, Colque, Crane,
  Damke, Davidson, Dean, Di~Mille, Don, Ebelke, Evans, Fitzgerald, Gillespie,
  Hall, Harding, Harding, Hammond, Hancock, Harrison, Hope, Horne, Karakla,
  Lam, Leger, MacDonald, Maseman, Matsunari, Melton, Mitcheltree, O’Brien,
  O’Connell, Patten, Richardson, Rieke, Rieke, Roman-Lopes, Schiavon, Sobeck,
  Stolberg, Stoll, Tembe, Trujillo, Uomoto, Vernieri, Walker, Weinberg, Young,
  Anthony-Brumfield, Bizyaev, Breslauer, Lee, Downey, Halverson, Huehnerhoff,
  Klaene, Leon, Long, Mahadevan, Malanushenko, Nguyen, Owen, Sánchez-Gallego,
  Sayres, Shane, Shectman, Shetrone, Skinner, Stauffer, \&
  Zhao}]{wilson_apache_2019}
Wilson, J.~C., Hearty, F.~R., Skrutskie, M.~F., {et~al.} 2019, Publications of
  the Astronomical Society of the Pacific, 131, 055001,
  \dodoi{10.1088/1538-3873/ab0075}

\bibitem[{Woods {et~al.}(2012)Woods, Ivanova, Van Der~Sluys, \&
  Chaichenets}]{woods_formation_2012}
Woods, T.~E., Ivanova, N., Van Der~Sluys, M.~V., \& Chaichenets, S. 2012, The
  Astrophysical Journal, 744, 12, \dodoi{10.1088/0004-637X/744/1/12}

\bibitem[{Zenati {et~al.}(2019)Zenati, Toonen, \&
  Perets}]{zenati_formation_2019}
Zenati, Y., Toonen, S., \& Perets, H.~B. 2019, Monthly Notices of the Royal
  Astronomical Society, 482, 1135, \dodoi{10.1093/mnras/sty2723}

\end{thebibliography}
\bibliographystyle{aasjournal}



\end{document}